\documentclass[pre,twocolumn,amsmath,amssymb,floatfix,superscriptaddress]{revtex4-1}
\usepackage{graphicx}
\usepackage{epstopdf}
\usepackage{amsmath}
\usepackage{amssymb}
\usepackage{color}
\usepackage{txfonts}
\usepackage{verbatim}
\usepackage{cprotect}
\usepackage{hyperref}

\begin{document}

\title{Impact of perception models on friendship paradox and opinion formation}

\author{Eun Lee}
\affiliation{Department of Mathematics, The University of North Carolina at Chapel Hill, Chapel Hill, North Carolina 27599, United States of America}
\author{Sungmin Lee}
\affiliation{Department of Physics, Korea University, Seoul 02841, Republic of Korea}
\author{Young-Ho Eom}
\affiliation{Department of Mathematics and Statistics, University of Strathclyde, Glasgow G1 1XH, United Kingdom}
\author{Petter Holme}
\affiliation{Institute of Innovative Research, Tokyo Institute of Technology, Yokohama, Kanagawa 226-8503, Japan}
\author{Hang-Hyun Jo}
\email[Corresponding author: ]{hang-hyun.jo@apctp.org}
\affiliation{Asia Pacific Center for Theoretical Physics, Pohang 37673, Republic of Korea}
\affiliation{Department of Physics, Pohang University of Science and Technology, Pohang 37673, Republic of Korea}
\affiliation{Department of Computer Science, Aalto University School of Science, Espoo FI-00076, Finland}
\date{\today}

\begin{abstract}
Topological heterogeneities of social networks have a strong impact on the individuals embedded in those networks. One of the interesting phenomena driven by such heterogeneities is the friendship paradox (FP), stating that the mean degree of one's neighbors is larger than the degree of oneself. Alternatively, one can use the median degree of neighbors as well as the fraction of neighbors having a higher degree than oneself. Each of these reflects on how people perceive their neighborhoods, i.e., their perception models, hence how they feel peer pressure. In our paper, we study the impact of perception models on the FP by comparing three versions of the perception model in networks generated with a given degree distribution and a tunable degree-degree correlation or assortativity. The increasing assortativity is expected to decrease network-level peer pressure, while we find a nontrivial behavior only for the mean-based perception model. By simulating opinion formation, in which the opinion adoption probability of an individual is given as a function of individual peer pressure, we find that it takes the longest time to reach consensus when individuals adopt the median-based perception model, compared to other versions. Our findings suggest that one needs to consider the proper perception model for better modeling human behaviors and social dynamics.
\end{abstract}

\maketitle

\section{Introduction}
\label{sec:introduction}

People's understanding of their social environment tends to be local~\cite{Galesic2012Social} and hence it can cause various biases~\cite{Feld1991Why, Lerman2016Majority, Lee2017Homophily}. In particular, the biases induced by a comparison of oneself with others, mostly his or her local neighbors, may have a number of consequences~\cite{Suls2002Social}, e.g., on self-esteem, subjective well being~\cite{Kross2013Facebook}, behavioral choices~\cite{Jackson2017Friendship, Hansen1991Preventing}, and the adoption of new technologies~\cite{Pinheiro2014Origin}. The social network is a substrate on which to build such biases. Recent studies on empirical social network datasets have revealed that the topological structure of social networks is highly heterogeneous, often characterized by degree heterogeneity~\cite{Barabasi1999Emergence, Broido2019Scalefree}, assortative mixing~\cite{Newman2002Assortative}, and community structure~\cite{Fortunato2010Community} to name a few. Such topological heterogeneities, partly summarized in Ref.~\cite{Jo2018Stylized}, can influence how people understand themselves and how they are influenced by others. This is important to understand various collective dynamics taking place in social networks, such as spreading, diffusion, and opinion formation~\cite{Pastor-Satorras2015Epidemic, Castellano2009Statistical, Sen2014Sociophysics}.

One of the most interesting phenomena driven by topological heterogeneities is the \emph{friendship paradox} (FP)~\cite{Feld1991Why}, sometimes called the ripple effect~\cite{Newman2003Egocentered}. The FP, stating that your friends have on average more friends than you do, has been explained by sampling bias~\cite{Feld1991Why, Eom2014Generalized, Jo2014Generalized}: Individuals with many friends tend to be observed more frequently by their friends. In many works, the individual's own degree has been compared to the mean degree of his or her neighbors~\cite{Ugander2011Anatomy, Eom2014Generalized, Jo2014Generalized, Momeni2016Qualities, Bollen2017Happiness, Iyer2018Friendship}. An alternative approach using the median instead of the mean has been taken~\cite{Ugander2011Anatomy, Kooti2014Network, Momeni2016Qualities}, which is often called the ``strong friendship paradox''~\cite{Kooti2014Network}. They used the median because it is less sensitive to neighbors with very high degrees~\cite{Lerman2016Majority, Wu2017NeighborNeighbor}. Although this topic has attracted attention, little is known about how such different approaches can affect people's perception of their neighborhoods as well as its consequences such as consensus time in opinion dynamics~\cite{Liggett1999Stochastic}.

In order to systematically study the impact of perception models on the friendship paradox and opinion formation, we compare three different versions of the perception model on how people integrate the information of their neighborhoods, namely, mean-based, median-based, and fraction-based versions. Here we newly introduce the fraction-based version in which the individual perceives his or her neighborhood in terms of the fraction of neighbors who have higher degrees than himself or herself. For studying the perception of individuals embedded in a network, we generate networks with a given degree distribution and a tunable degree-degree correlation or assortativity. One can naturally expect that the more assortative network may weaken FP for individuals as they become surrounded by similar others. Therefore, by adopting the network model with a tunable assortativity, we can better understand the interplay between the perception model and network structure. Based on the individual perceptions using different versions of the perception model, we derive degree-dependent and network-level perceptions as well as their effects on the opinion dynamics. In particular, we focus on how the consensus time of a nonequilibrium opinion formation model can be affected by the different versions of the perception model.

Our paper is organized as follows: We introduce our model in Sec.~\ref{sec:model}. In Sec.~\ref{sec:results}, by numerical simulations, we compare different perception models in terms of the friendship paradox and opinion formation with the exponential degree distribution. The other degree distributions, i.e., binomial and power-law degree distributions, are tested in Sec.~\ref{sec:other}. We finally conclude our work in Sec.~\ref{sec:conclusion}.

\section{Model}
\label{sec:model}

\subsection{Neighborhood perception and peer pressure}
\label{subsec:perception}

In order to study how each individual embedded in a network perceives his or her neighborhood, let us consider the case where each individual (ego) compares his or her own degree to the degrees of his or her neighbors. For this comparison, we use three different versions of the perception model reflecting on how the perceptions of neighbors are integrated, in terms of (i) the mean degree of neighbors, (ii) the median degree of neighbors, and (iii) the fraction of neighbors having a higher degree than the ego.

\begin{figure}[!t]
    \includegraphics[width=0.9\columnwidth]{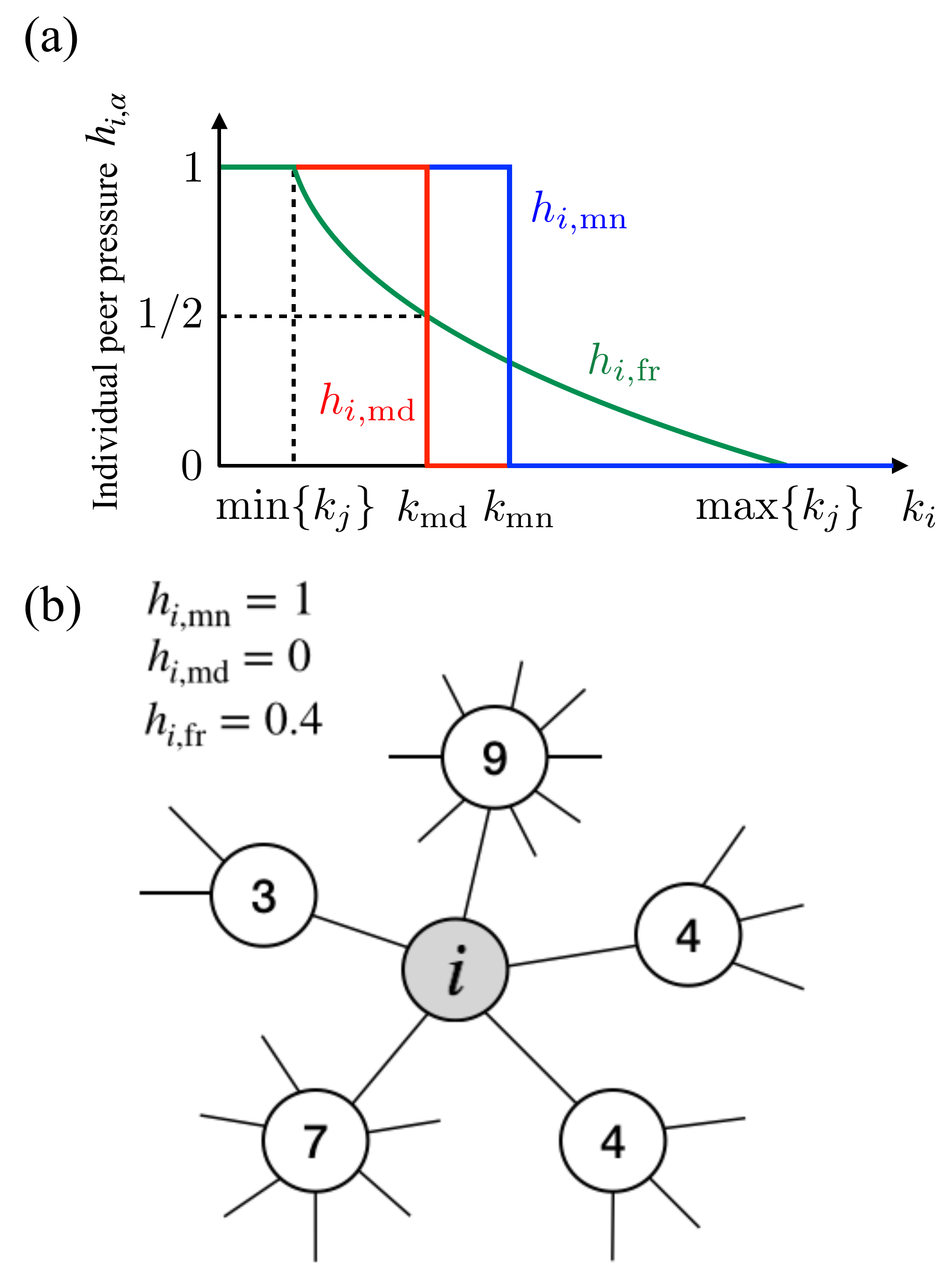}
    \caption{Schematic diagram of the three versions of the neighborhood perception; the mean-based, mean-based, and fraction-based ones, denoted by $h_{i,\rm mn}$ in Eq.~\eqref{eq:h_mn}, $h_{i,\rm md}$ in Eq.~\eqref{eq:h_md}, and $h_{i,\rm fr}$ in Eq.~\eqref{eq:h_fr}, respectively. In panel (a), $k_i$ denotes the degree of the node $i$ and $k_j$ does the degree of $i$'s neighbor $j$. $k_{\rm mn}$ and $k_{\rm md}$ are the mean and median degrees of $i$'s neighbors. It is also possible that $k_{\rm md}$ is larger than $k_{\rm mn}$ depending on the degrees of neighbors. Panel (b) shows an example of the differences of our three versions of the neighborhood perception. The number in each surrounding circle represents the degree of the node.}
    \label{fig1:schematic}
\end{figure}

As for the mean-based version, if the ego's degree is smaller than the mean degree of its neighbors, then we say that the ego experiences peer pressure. Although in reality, this may vary with the ego's personality, for simplifying the discussion we can say that people feel pressure when they are different from their perception of a typical other in the population. Precisely, for a node $i$, we define the \emph{mean-based peer pressure} as follows:
\begin{equation}
    \label{eq:h_mn}
    h_{i,\rm mn}\equiv \theta\left(\frac{1}{k_i}\sum_{j\in \Lambda_i} k_j - k_i\right),
\end{equation}
where $\Lambda_i$ denotes a set of $i$'s neighbors and $k_i\equiv |\Lambda_i|$ is the degree of the node $i$. Note that as $\theta(\cdot)$ is the Heaviside step function, the above mean-based peer pressure $h_{i,\rm mn}$ is a binary variable having the value of either $0$ or $1$, which can be related to the paradox holding probability defined in Refs.~\cite{Eom2014Generalized, Jo2014Generalized}. Next we introduce the \emph{fraction-based peer pressure}, whose value is defined as the fraction of neighbors having a higher degree than the ego:
\begin{equation}
    \label{eq:h_fr}
    h_{i,\rm fr}\equiv \frac{1}{k_i}\sum_{j\in \Lambda_i}\theta(k_j -k_i).
\end{equation}
Finally, the \emph{median-based peer pressure} can be defined directly from the fraction-based peer pressure as
\begin{equation}
    \label{eq:h_md}
    h_{i,\rm md} \equiv \theta\left(h_{i,\rm fr} - \frac{1}{2}\right).
\end{equation}
Note that $h_{i,\rm fr}$ is a real number in $[0,1]$, while $h_{i,\rm md}$ is a binary variable. The median-based peer pressure has been related to the strong form of the friendship paradox~\cite{Kooti2014Network}. The different behaviors between these three versions are depicted in Fig.~\ref{fig1:schematic}, where $k_{\rm mn}$ and $k_{\rm md}$ denote the mean degree and median degree of neighbors of the ego $i$, respectively. We like to note that these three versions could be seen as special cases of a general formula, which is briefly discussed in Appendix~\ref{append:general}.

These definitions of individual peer pressure are applicable to both degree-dependent peer pressure as well as the network-level peer pressure. The \emph{degree-dependent peer pressure} is defined as the average peer pressure for the set of nodes with the same degree:
\begin{equation}
    \label{eq:h_degree}
    h_{\alpha}(k) \equiv \langle h_{i,\alpha}\rangle_{\{i|k_i=k\}},
\end{equation}
where $\alpha\in \{{\rm mn, md, fr}\}$ denotes the respective version. The degree-dependent approach, assuming that all nodes of the same degree are statistically equivalent, has been extensively studied in network science~\cite{Pastor-Satorras2015Epidemic}. The network-level peer pressure for a network of size $N$ is as follows:
\begin{equation}
    \label{eq:h_network}
    H_{\alpha} \equiv \dfrac{1}{N}\sum_{i=1}^N h_{i,\alpha}.
\end{equation}
If the degree distribution of the network is right-skewed as in many empirical social networks~\cite{Broido2019Scalefree}, then the degree distribution for neighbors of a node $i$ would also be typically right-skewed. This implies that $k_{\rm md}<k_{\rm mn}$ for the node $i$, hence $h_{i,\rm md}\leq h_{i,\rm mn}$, as depicted in Fig.~\ref{fig1:schematic}(a). If this inequality holds for a majority of nodes, then it might also be the case with the network-level peer pressures: $H_{\rm md}\lesssim H_{\rm mn}$.

\subsection{Network generation}
\label{subsec:network}

In order to study the effects of our three models of how individuals perceive their neighborhoods, we generate networks with a given degree distribution and a tunable degree-degree correlation. For this, we adopt the procedure suggested in Ref.~\cite{Jo2014Generalized}: We first construct an uncorrelated network with a given degree distribution using the configuration model~\cite{Catanzaro2005Generation}, and then 
we rewire links until a desired degree-degree correlation is reached~\cite{Holme2007Exploring} (see further discussion below).

For the construction of an uncorrelated network, we generate a degree sequence $\{k_i\}$ for nodes $i=1,\cdots,N$. Here each degree is independently drawn from the given degree distribution $P(k)$. The node $i$ is given by $k_i$ stubs or half links. We randomly choose a pair of nodes and create a link between them if there is no link between them and if both nodes have residual stubs. This linking process is repeated until when all stubs are used up. 

To estimate the degree-degree correlation, we first define the assortativity coefficient for a network with $L$ links~\cite{Newman2002Assortative}:
\begin{equation}
    \label{eq:assort}
    r_{kk} \equiv \frac{L\sum_l k_lk'_l- \left[\sum_l\frac{1}{2}(k_l+k'_l)\right]^2} {L\sum_l\frac{1}{2}({k_l}^2+{k'_l}^2) -\left[\sum_l\frac{1}{2}(k_l+k'_l)\right]^2},
\end{equation}
where $k_l$ and $k'_l$ denote degrees of ending nodes of the $l$th link with $l=1,\cdots,L$. $r_{kk}$ measures the tendency that high-degree nodes are connected to each other and so are low-degree nodes. The value of $r_{kk}$ can range from $-1$ for the extremely disassortative case to $1$ for the extremely assortative case. The rewiring process is as follows~\cite{Maslov2002Specificity}: We randomly choose two links, say, $(i,j)$ and $(i',j')$. These two links are cut and replaced by either $(i,i')$ and $(j,j')$, or $(i,j')$ and $(i',j)$, only when the rewiring makes $r_{kk}$ closer to the desired value. This rewiring process is repeated until the desired value of $r_{kk}$ is reached. By this method, the degree distribution remains the same irrespective of the degree-degree correlation~\footnote{For the uncorrelated network to be consistent with networks with other values of $r_{kk}$, we first generate slightly negatively correlated networks with $r_{kk}=-0.05$ for the exponential and binomial $P(k)$ and $r_{kk}=-0.01$ for the power-law $P(k)$, and then reapply the rewiring to make the networks with $r_{kk}=0$.}.

\section{Results}
\label{sec:results}

\subsection{Peer pressures in degree-correlated networks}
\label{subsec:peer}

\begin{figure}[!t]
    \includegraphics[width=0.95\columnwidth]{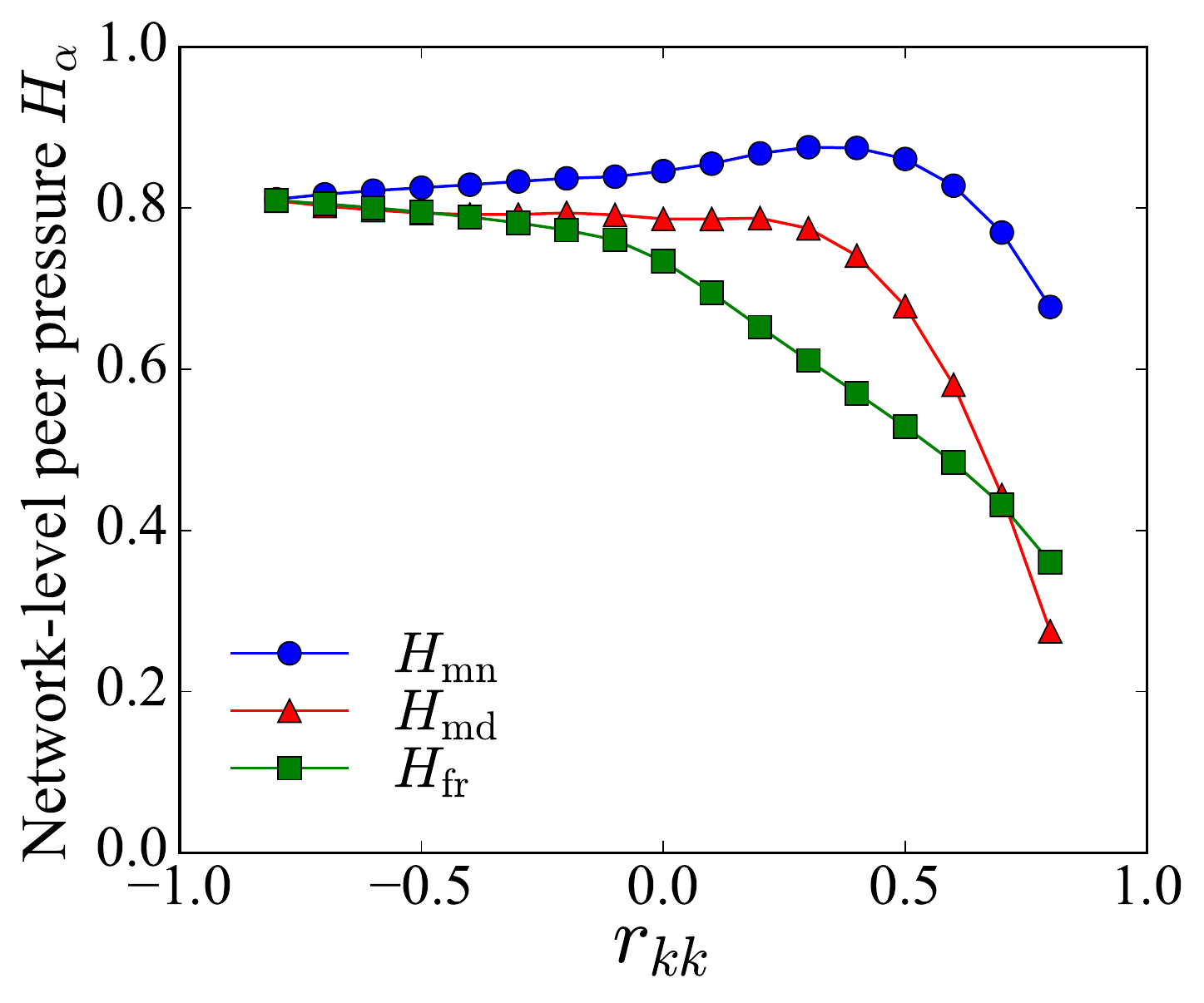}
    \caption{Network-level peer pressures $H_{\alpha}$ for $\alpha\in\{{\rm mn, md, fr}\}$ in Eq.~\eqref{eq:h_network} as a function of $r_{kk}$, when $P(k)$ in Eq.~\eqref{eq:Pk_exp} is used. For each value of $r_{kk}$, we generated $50$ different networks of size $N=5\times 10^4$ to obtain the average value of $H_{\alpha}$. The standard errors are smaller than the symbols.}
    \label{fig:H}
\end{figure}

In this section, we study the case with an exponential degree distribution
\begin{equation}
    \label{eq:Pk_exp}
    P(k)=\langle k\rangle^{-1} e^{-k/\langle k\rangle},
\end{equation}
with the average degree of $\langle k\rangle=50$ to generate networks with various values of $r_{kk}$. Then one can measure the peer pressures at different levels as discussed in Sec.~\ref{subsec:perception}. We begin with the network-level peer pressures $H_{\alpha}$ for $\alpha\in \{{\rm mn, md, fr}\}$, which clearly shows the importance of the perception model, as depicted in Fig.~\ref{fig:H}. We immediately observe that $H_{\rm md}\leq H_{\rm mn}$ for the entire range of $r_{kk}$, possibly due to the right-skewed distribution of $P(k)$ in Eq.~\eqref{eq:Pk_exp}.

\begin{figure*}[!t]
   \includegraphics[width=\textwidth]{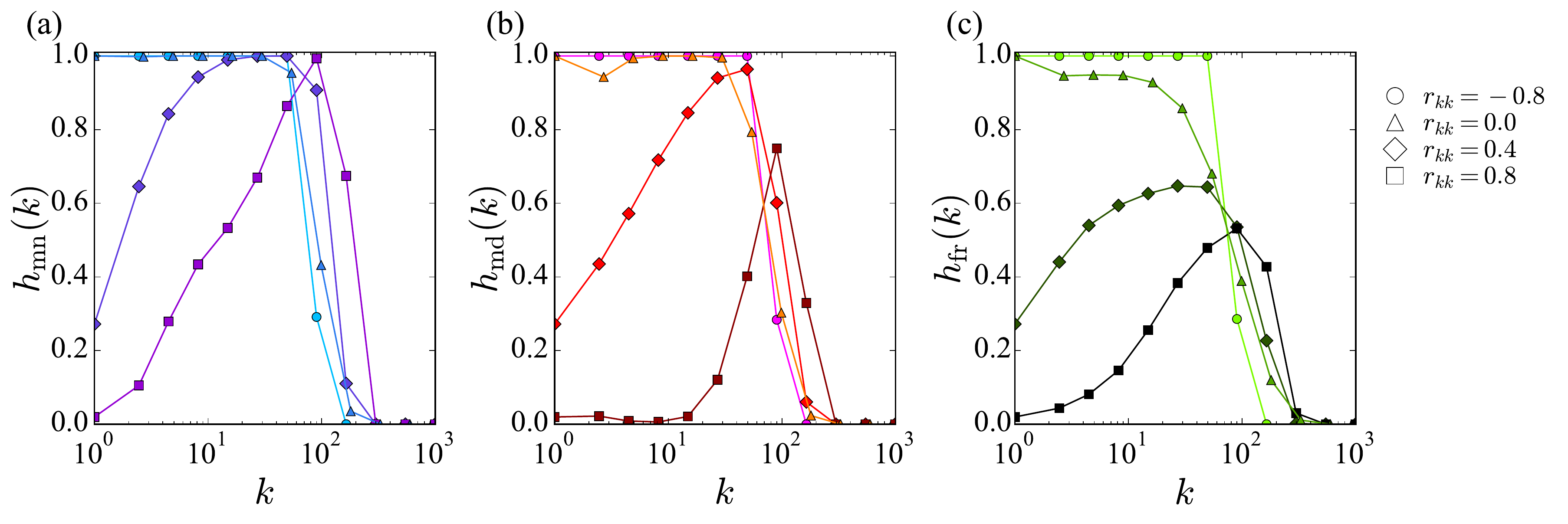}
   \caption{Degree-dependent peer pressures $h_{\alpha}(k)$ for $\alpha\in\{{\rm mn, md, fr}\}$ in Eq.~\eqref{eq:h_degree} for several values of $r_{kk}$, derived from the same networks used in Fig.~\ref{fig:H}.}
    \label{fig:hk}
\end{figure*}

When the network is largely disassortative, e.g., when $r_{kk}\approx -0.8$, all three $H_{\alpha}$ have almost the same value. In such a disassortative network the hub nodes tend to be separated while being connected to a number of low-degree nodes. Then a majority of nodes with low degrees would have the similar levels of peer pressure irrespective of the perception models. 

This picture is consistent with the step-shaped curves of degree-dependent peer pressures $h_{\alpha}(k)$ for all $\alpha\in \{{\rm mn, md, fr}\}$, as shown in Fig.~\ref{fig:hk}. As the value of $r_{kk}$ increases from $-0.8$ to $0$, we find that $H_{\rm fr}$ slightly decreases and $H_{\rm md}$ remains almost the same, while only $H_{\rm mn}$ slightly increases. In other words, as the network becomes less disassortative, some high-degree nodes begin to feel the peer pressure as they are rewired from low-degree nodes to nodes with even higher degrees than their own degrees, which can be called an \emph{escalation effect}. At the same time, some low-degree nodes are no longer under peer pressure by losing connections to high-degree nodes and by being connected to nodes with even lower degrees, which can be called a \emph{relaxation effect}. Although both effects are observed for all versions of the perception model as shown in Fig.~\ref{fig:hk}, the dominant effect will decide the overall trend of the curve of $H_{\alpha}$. For example, in the median-based case, two effects appear to be balanced so that $H_{\rm md}$ could remain almost the same for the range of $r_{kk}<0$. In addition, we derive $H_{\rm fr}=3/4$ for $r_{kk}=0$ as detailed in the Appendix~\ref{append:Hfr0}, comparable to the numerical result $0.73 \pm 0.04$ in Fig.~\ref{fig:H}. 

In the case with positive $r_{kk}$, i.e., when networks are assortative, we find that $H_{\rm fr}$ monotonically decreases according to $r_{kk}$, while $H_{\rm md}$ remains almost constant for $r_{kk}<0.3$, and then it starts to sharply decrease. Interestingly, $H_{\rm mn}$ keeps increasing for $r_{kk}<0.4$ before decreasing, but its value at $r_{kk}=0.8$ is still much higher than those of $H_{\rm md}$ and $H_{\rm fr}$. In general, such overall decreasing behaviors can be understood as nodes tend to be connected to other nodes with similar degrees, suppressing peer pressure. However, this tendency for assortativity depends on the degree: The low-degree nodes show most drastic changes in their peer pressures because even a small number of assortative rewiring can make those nodes have a strong relaxation effect. On the other hand the high-degree nodes seem to be most robust with respect to the rewiring as they still have many neighbors with small degrees. The nodes with intermediate degrees typically experience the highest peer pressure as they are still connected to a number of nodes with higher degrees. 

In these cases, it turns out that the role of perception models become crucial, e.g., as depicted by the degree-dependent peer pressures for $r_{kk}=0.4$ in Fig.~\ref{fig:hk}. The mean-based version leads to the highest level of peer pressure for the range of $10<k<100$, and then the median-based one follows. The fraction-based version shows the lowest level of peer pressure in the same range of degree. It is because the mean-based version is more sensitive to the connections to the high-degree nodes than other versions. Such a relative high level of peer pressure in $h_{\rm mn}(k)$ for $10<k<100$, combined with the large fraction of nodes in the same range of degree, must be the reason why $H_{\rm mn}$ shows the increasing behavior for the range of $0<r_{kk}<0.4$. As $r_{kk}$ becomes larger than $0.4$, even nodes with intermediate degrees start to have the relaxation effect, leading to the decreasing behavior of $H_{\rm mn}$. Finally, low levels of $H_{\rm md}$ and $H_{\rm fr}$ for $r_{kk}>0.6$ can be understood mostly by the small values of degree-dependent peer pressures for low-degree nodes, as shown in Figs.~\ref{fig:hk}(b) and~\ref{fig:hk}(c).

Therefore, in order to reduce the network-level peer pressure in a society of individuals adopting the mean-based perception model, decreasing the assortativity can also be a feasible solution, while it can end up with a relatively high level of peer pressure even in the extremely disassortative case.

\subsection{Opinion formation affected by perception models}
\label{subsec:opinion}

Since people take actions based on their perceptions of their social environment, the different versions of the perception model can affect collective dynamics taking place on social networks. We will address this issue by introducing peer-pressure effects on the stochastic voter model.

\begin{figure}[!t]
\centering
    \includegraphics[width=0.9\linewidth]{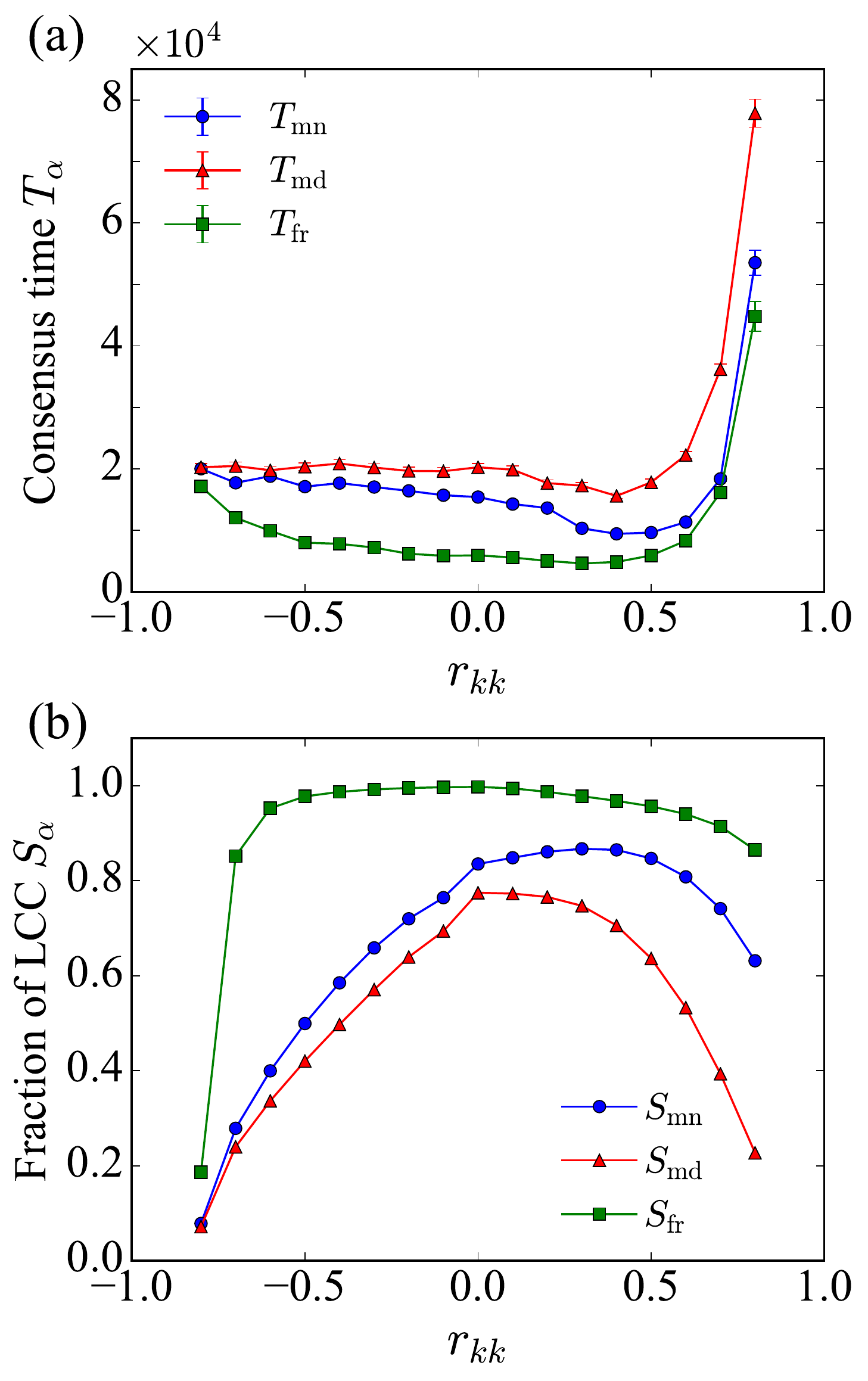}
    \caption{(a) Consensus times, $T_{\alpha}$, in the unit of Monte Carlo time steps for our voter model and (b) fractions of the largest connected components (LCCs) in the effective networks, $S_{\alpha}$, for $\alpha\in\{{\rm mn, md, fr}\}$ as a function of $r_{kk}$. The voter model was simulated on $100$ different networks of size $N=10^4$ and for $P(k)$ in Eq.~\eqref{eq:Pk_exp}, with $10$ random initial conditions for each network, to obtain the average consensus times. The fractions of LCCs were obtained from the corresponding networks of the simulations. The standard errors are shown in (a), but not in (b) as they are smaller than the size of symbols.}
    \label{fig:consensus}
\end{figure}

In a conventional deterministic voter model defined on a network, e.g., in Ref.~\cite{Sood2008Voter}, each node can have an opinion of either $1$ or $-1$. Initially, some fraction of nodes, denoted by $\rho$, have the opinion of $1$, while the others have the opinion of $-1$. At each time step, one node, say $i$, is randomly chosen, and one of $i$'s neighbors, say $j$, is also randomly chosen. Then the node $i$ replaces its opinion by $j$'s opinion with the adoption probability $1$. This adoption process is repeated until all nodes have the same opinion, implying the consensus. The time spent to reach the consensus is called the consensus time, denoted by $T$. 

We extend this voter model by introducing the adoption probability for each node $i$ as a function of its peer pressure, precisely, as follows:
\begin{equation}
    p_{i,\alpha} \equiv \epsilon + (1-\epsilon)h_{i,\alpha},
    \label{eq:adoption}
\end{equation}
where $\epsilon$ is a small positive number and it denotes a base adoption probability. $h_{i,\alpha}$ is the peer pressure of the node $i$ when the respective version $\alpha\in\{{\rm mn, md, fr}\}$ is used. If $h_{i,\alpha}=1$, then one gets $p_{i,\alpha}=1$, while $p_{i,\alpha}=\epsilon$ when $h_{i,\alpha}=0$. Here we have assumed that the larger peer pressure of a node may lead to the larger probability of adopting his or her neighbor's opinion~\cite{Jackson2017Friendship}. The small positive $\epsilon$ is introduced to avoid the case with diverging consensus times due to some individuals who never change their opinion. Note that the adoption probability is solely determined by the peer pressure derived from the local network topology, not by the opinions of the individuals.

Considering the average adoption probability given by $\langle p_{i,\alpha}\rangle \equiv \frac{1}{N}\sum_i p_{i,\alpha} = \epsilon + (1-\epsilon)H_{\alpha}$, one can naturally expect that the consensus times $T_{\alpha}$ for $\alpha\in\{{\rm mn, md, fr}\}$ will be negatively correlated with the network-level peer pressures $H_{\alpha}$. For testing this expectation, we simulate our voter model with $\rho=0.2$ and $\epsilon=0.05$ on the networks generated using the method in Sec.~\ref{subsec:network}. In Fig.~\ref{fig:consensus}(a), we numerically find the overall opposite trend of $T_{\rm mn}$ ($T_{\rm md}$) compared to $H_{\rm mn}$ ($H_{\rm md}$), see also Fig.~\ref{fig:H}: The voter model using the mean-based peer pressure results in a faster consensus than that of the median-based one. However, in the fraction-based case, both $T_{\rm fr}$ and $H_{\rm fr}$ are the lowest for the almost entire range of $r_{kk}$, compared to other versions of the perception model. Moreover, we also find a nontrivial pattern of $T_{\rm fr}$ that it decreases for the range of $r_{kk}<0$ although $H_{\rm fr}$ monotonically decreases in the entire range of $r_{kk}$. 

Such unexpected results can be partly explained by the nonbinary nature of the individual peer pressure $h_{i,\rm fr}$ by definition in Eq.~\eqref{eq:h_fr}: For the mean-based and median-based cases, the individual peer pressure has a value of either $0$ or $1$, leading to the adoption probability of $\epsilon$ or $1$, respectively. Then, nodes with the adoption probability of $\epsilon$ play a role of bottlenecks on the opinion formation, generically increasing the consensus times. In contrast, the real-valued peer pressures in the fraction-based case tend to accelerate the opinion formation, hence the lowest level of $T_{\rm fr}$ is observed. The nonbinary effect can account for the decreasing $T_{\rm fr}$ in the range of $r_{kk}<0$ as well. With the extremely negative $r_{kk}$, the individual fraction-based peer pressures yet have binary nature. As $r_{kk}$ increases to $0$, the binary nature weakens to accelerate the opinion formation process, leading to the decreasing $T_{\rm fr}$ in spite of the slight decrease of $H_{\rm fr}$.

For a more detailed understanding of the consensus times, we study the bottleneck effects due to nodes with very low peer pressures on the opinion formation. For this, an effective network is derived from an original network by keeping only nodes of $h_{i,\alpha}>0$ and links between them, while removing nodes of $h_{i,\alpha}=0$ from the network. In general, the effective network may consist of more than one connected component (CC). The consensus times can be inferred not only by the size distribution of CCs, but also by how they are connected by the nodes with zero peer pressure. We first measure the fraction of the largest connected component (LCC), denoted by $S_{\alpha}$, as it has been known to strongly affect the consensus times in the previous work~\cite{Lee2017Modeling}. In Fig.~\ref{fig:consensus}(b), we indeed find that the behavior of $S_{\alpha}$ for different versions of the perception model can explain the consensus times $T_{\alpha}$ more consistently than $H_{\alpha}$.

However, we also find somewhat atypical behaviors. In particular, we focus on two extreme cases of $r_{kk}$: On the one hand, when $r_{kk} \approx 0.8$, the values of $S_{\alpha}$ for $\alpha\in\{{\rm mn, md, fr}\}$ are not the lowest, whereas all $T_{\alpha}$s have the highest values. In this case, in addition to the LCC, we find several relatively big CCs (not shown). If the nodes in some of such CCs quickly reach consensus within CCs but with the opposite opinion to that of the LCC, then it will take longer times to revert such local consensus for reaching the global consensus. On the other hand, if $r_{kk} \approx -0.8$, then the values of $S_{\alpha}$ are the lowest, whereas $T_{\alpha}$ has the relatively low values. As the networks can be described by the starlike structure in this case, their effective networks, i.e., without hub nodes with $h_{i,\alpha}=0$, are found to consist of the LCC of relatively small size and a number of isolated nodes (not shown). In contrast to the case with $r_{kk}\approx 0.8$, these isolated nodes are more vulnerable than the connected nodes to the opinions of hub nodes. Thus, once the hub nodes reach consensus via dangling nodes connecting hub nodes, the global consensus will be formed in a short time period. More detailed understanding of the opinion formation process is left for future works.

\begin{figure}[!t]
\centering
    \includegraphics[width=\linewidth]{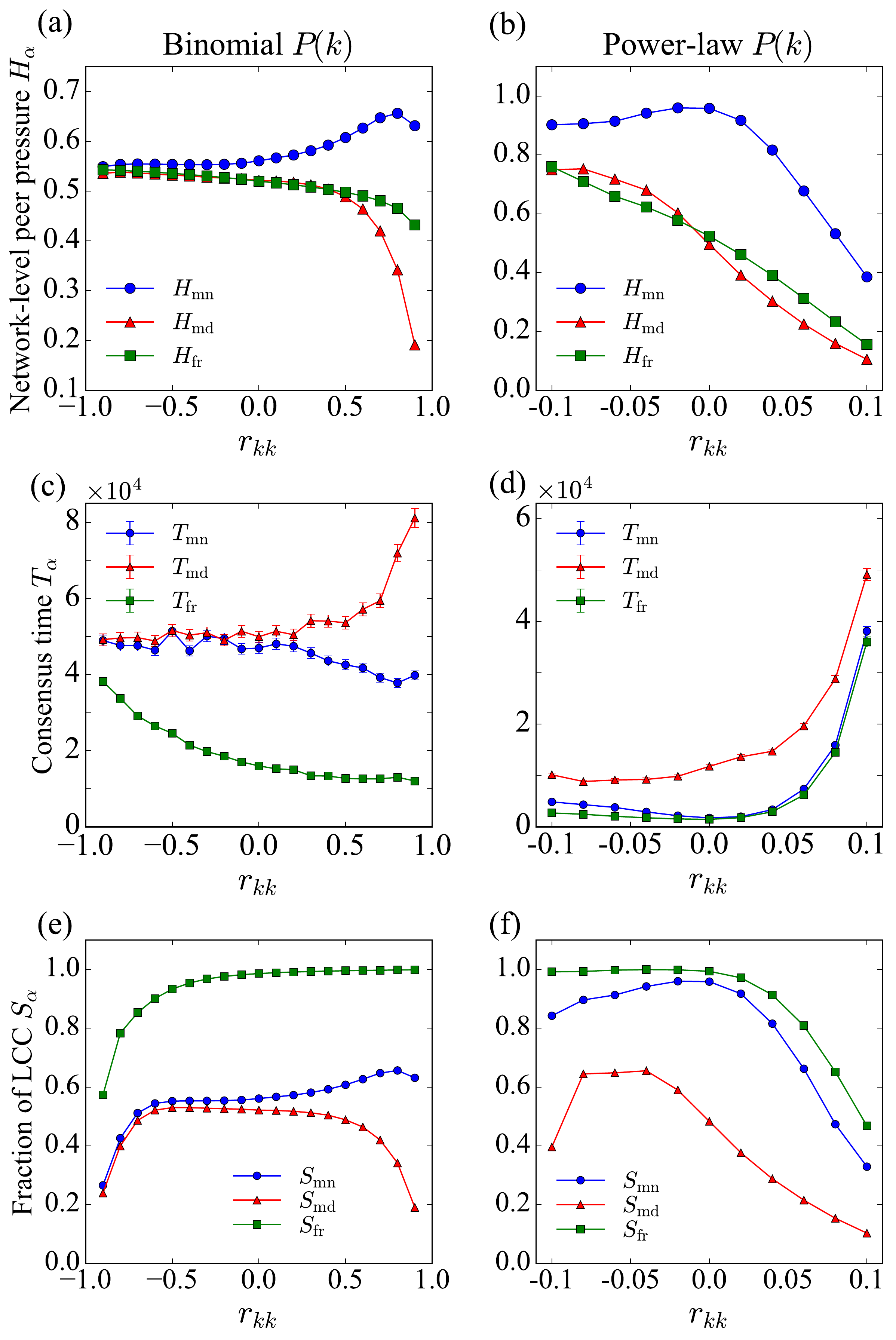}
    \caption{Effects of perception models and network structures on network-level peer pressures $H_{\alpha}$ (top), consensus times $T_{\alpha}$ (middle), and fractions of LCCs (bottom), as functions of $r_{kk}$ for $\alpha \in \{\rm mn, md, fr\}$. A binomial degree distribution (left) and power-law degree distribution (right) are tested on $100$ different networks of size $N=10^4$. Standard errors are shown when the error is larger than the size of symbols.}
    \label{fig:othernet}
\end{figure}

\section{Other degree distributions}
\label{sec:other}

In order to test the effects of the degree distribution on the statistical properties of peer pressures as well as opinion formation, we adopt two other degree distributions, i.e., binomial distribution and power-law distribution. The binomial degree distribution can be obtained for Erd\"{o}s-R\'{e}nyi random graphs with linking probability $p$~\cite{Erdos1960Evolution}:
\begin{equation}
	P(k)={{N-1} \choose k}p^k(1-p)^{N-1-k}.
\end{equation}
In our work, we choose the value of $p$ leading to $\langle k\rangle=50$ as in Sec.~\ref{sec:results}. The power-law degree distribution is a key feature of various scale-free networks~\cite{Barabasi1999Emergence, Albert1999Diameter, Clauset2009Powerlaw}. Here we use the following form:
\begin{equation}
    P(k)\propto k^{-\gamma}\ \textrm{for}\ k\geq k_{\rm min},
\end{equation}
with $\gamma=2.7$ and $k_{\rm min}=6$, leading to $\langle k\rangle\approx 13$. Once the degree sequence is constructed by drawing $N$ random values from a given $P(k)$, we apply the same method described in Sec.~\ref{subsec:network} to generate the networks with a desired value of assortativity $r_{kk}$. We note that the range of $r_{kk}$ is strongly limited by the shape of $P(k)$~\cite{Menche2010Asymptotic}: We study the range of $-0.1\leq r_{kk}\leq 0.1$ in the case with power-law degree distributions.

We measure the network-level peer pressures $H_{\alpha}$ as shown in Figs.~\ref{fig:othernet}(a) and~\ref{fig:othernet}(b), to find similar behaviors of $H_{\alpha}$ to those in the case with the exponential degree distribution in Sec.~\ref{sec:results}. In both binomial and power-law cases, as $r_{kk}$ increases, $H_{\rm mn}$ shows increasing and then decreasing behaviors, while $H_{\rm md}$ and $H_{\rm fr}$ monotonically decrease. Overall the curve of $H_{\rm mn}$ is the highest compared to $H_{\rm md}$ and $H_{\rm fr}$. Then, we simulate our voter model using $\rho=0.2$ and $\epsilon=0.05$ on these networks to measure the consensus times, depicted in Figs.~\ref{fig:othernet}(c) and~\ref{fig:othernet}(d). We find the opposite trends between $T_{\alpha}$ and $H_{\alpha}$ for the mean-based and median-based cases, while the curves of $T_{\rm fr}$ appear to be the lowest compared to the other versions. This is again partly due to the nonbinary nature of the fraction-based peer pressure. Also, the trend of $T_{\alpha}$ can be explained in terms of the fraction of LCC $S_{\alpha}$ of their corresponding effective networks, as shown in Figs.~\ref{fig:othernet}(e) and~\ref{fig:othernet}(f). We like to note that other parameter values---$\langle k\rangle=30$ for exponential and binomial $P(k)$ and $\gamma=3.5$ for power-law $P(k)$---do not change our conclusions (not shown).

\section{Conclusion}
\label{sec:conclusion}

In order to study the impact of the perception models on the friendship paradox and opinion formation, we have compared three versions of the perception model reflecting on how people perceive their neighborhoods, which are mean based, median based, and fraction based, respectively. These three versions of the perception model are tested in networks with a given degree distributions and a tunable degree-degree correlation. By the numerical simulations, we find that the perception models applied to individuals embedded in a network indeed affect the network-level peer pressure as well as the collective dynamics in terms of the consensus times of the opinion formation. 

Regarding network-level peer pressure, one can naturally expect that the more assortative network may lead to the lower peer pressure at the network level, which is indeed the case with the median-based and fraction-based cases. However, the mean-based peer pressure at the network level increases and then decreases as the assortativity changes from $-0.8$ to $0.8$ in the case with the exponential degree distribution, which is in contrast to the naive expectation. This unexpected result can be understood by the finer observations using the degree-dependent peer pressures. When starting from the disassortative network, the assortative rewirings have asymmetric effects depending on the degree of nodes: The low-degree (high-degree) nodes are more vulnerable (robust) to such rewirings, hence showing more (less) drastic changes in their peer pressures. The intermediate-degree nodes, comprising a considerable portion of the population, tend to have the highest peer pressures in assortative networks for all versions of the perception model. However, their peer pressures are more sensitive to the structural change when the mean-based perception model is used, leading to the increasing behavior of mean-based peer pressure at the network level.

The perception models also turn out to have a strong impact on the consensus times of our voter model where the adoption probability of the individual is given as a function of individual peer pressure. We find that the median-based perception model results in the longest consensus time, while the shortest consensus times are found in the fraction-based case. It means that the network-level peer pressures cannot fully account for the consensus times. For this, we derive the effective network from the original network by removing the nodes with zero peer pressure that are a bottleneck in opinion formation. We find that the behavior of the largest connected component (LCC) and other connected components (CCs) can to a large extent explain the observed results. Again the different versions of the perception model result in the different size distributions of the CCs. While the size of the LCC can be used to predict the consensus times more consistently than the network-level peer pressure, the longest consensus times found in the extremely assortative networks are possibly due to the longer time it takes to revert the local consensus with the opposite opinion to the global consensus in the sizable CCs. 

As demonstrated in our work, not only the network structure but also the perception models of individuals can have a crucial impact on how people understand themselves and how they are influenced by others. Since such perception process and its consequences are largely unexplored, experimental tests for clarifying the perception process and measuring peer pressure can be conducted as to expand our understanding in this field. Alternatively, we can take various analytical and numerical approaches to better understand the individual perceptions and their impact on the network-level properties and various collective dynamics taking place in social networks, as well as eventually the coevolution of individual perceptions and network structure~\cite{Jo2016Dynamical}.

\begin{acknowledgments}
S.L. was supported by Basic Science Research Program through the National Research Foundation of Korea (NRF) funded by the Ministry of Education (2016R1A6A3A11932833). Y.-H.E. acknowledges support from the University of Strathclyde and funding from Ministerio de Econom\'{\i}a y Competividad (Spain) through project FIS2016-78904-C3-3-P. P.H. was supported by JSPS KAKENHI Grant Number JP 18H01655. H.-H.J. acknowledges financial support by Basic Science Research Program through the National Research Foundation of Korea (NRF) grant funded by the Ministry of Education (NRF-2018R1D1A1A09081919). 
\end{acknowledgments}

\appendix

\section{General formula for the perception model}
\label{append:general}

The three versions of the perception of the neighborhood defined in Eqs.~\eqref{eq:h_mn}--\eqref{eq:h_md} can be generalized to the following formula:
\begin{equation}
    \label{eq:h_general}
    h_i \equiv f\left[\frac{1}{k_i}\sum_{j\in \Lambda_i}g(k_j -k_i)-\phi_i \right],
\end{equation}
where $\phi_i$ denotes some characteristic value of the node $i$. The mean-based peer pressure is recovered if $f(x)=\theta(x)$, $g(x)=x$, and $\phi_i=0$. The median-based peer pressure is recovered if $f(x)=\theta(x)$, $g(x)=\theta(x)$, and $\phi_i=1/2$. Finally, the fraction-based peer pressure is recovered if $f(x)=x$, $g(x)=\theta(x)$, and $\phi_i=0$. Therefore, one can explore other versions of the perception model by specifying $f(x)$, $g(x)$, and $\phi_i$ in the above general formula.

\section{Derivation of $H_{\rm fr}=3/4$ in the case with $r_{kk}=0$}
\label{append:Hfr0}

Although the analytical calculation of the network-level peer pressure in the correlated network is not straightforward in general, some special cases can be analyzed. Here we derive $H_{\rm fr}$ in the case with the uncorrelated network, i.e., $r_{kk}=0$. From Eqs.~\eqref{eq:h_fr} and~\eqref{eq:h_degree}, the degree-dependent fraction-based peer pressure can be written as
\begin{equation}
    \label{eq:h_fr_k_approx}
    h_{\rm fr}(k)= \frac{1}{k}\sum_{j=1}^k \theta(k_j - k)\approx \int_k^\infty P(k'|k)dk'.
\end{equation}
Since $r_{kk}=0$, we have the conditional degree distribution 
\begin{equation}
P(k'|k)=\frac{k'P(k')}{\langle k\rangle}.
\end{equation}
Then by using the exponential degree distribution in Eq.~\eqref{eq:Pk_exp}, one gets
\begin{equation}
    h_{\rm fr}(k)= \left(1+\frac{k}{\langle k\rangle}\right)e^{-k/\langle k\rangle},
\end{equation}
from which we calculate the network-level peer pressure as
\begin{equation}
    H_{\rm fr}= \int_0^\infty h_{\rm fr}(k)P(k)dk=\frac{3}{4}.
    \label{eq:Hfr_r0}
\end{equation}

\bibliographystyle{apsrev4-1}
%\bibliography{/Users/h2jo/Research/_papers/h2jo-papers}
%merlin.mbs apsrev4-1.bst 2010-07-25 4.21a (PWD, AO, DPC) hacked
%Control: key (0)
%Control: author (72) initials jnrlst
%Control: editor formatted (1) identically to author
%Control: production of article title (-1) disabled
%Control: page (0) single
%Control: year (1) truncated
%Control: production of eprint (0) enabled
%

\end{document}